# Data-driven dose calculation algorithm based on deep learning

Jiawei Fan[1,†], Lei Xing[1,*], Peng Dong[1], Jiazhou Wang[2], Weigang Hu[2], and Yong Yang[1,*]

1. Department of Radiation Oncology, Stanford University, 875 Blake Wilbur Drive, Stanford, CA 94305-5847, USA

2. Department of Radiation Oncology, Fudan University Shanghai Cancer Center; Department of Oncology, Shanghai Medical College, Fudan University, Shanghai 200032, China

* Corresponding authors:
lei@Stanford.edu
yong66@stanford.edu

† On leave from Department of Radiation Oncology, Fudan University Shanghai Cancer Center; Department of Oncology, Shanghai Medical College, Fudan University, Shanghai 200032, China

**Purpose:** Accurate and efficient dose calculation is an important prerequisite to ensure the success of radiation therapy. However, all the dose calculation algorithms commonly used in current clinical practice have to compromise between calculation accuracy and efficiency, which may result in unsatisfactory dose accuracy or highly intensive computation time in many clinical situations. The purpose of this work is to develop a novel dose calculation algorithm based on the deep learning method for radiation therapy.

**Methods:** In this study we performed a feasibility investigation on implementing a fast and accurate dose calculation based on a deep learning technique. A two dimensional (2D) fluence map was first converted into a three dimensional (3D) volume using ray traversal algorithm. A 3D

U-Net like deep residual network was then established to learn a mapping between this converted 3D volume, CT and 3D dose distribution. Therefore an indirect relationship was built between a fluence map and its corresponding 3D dose distribution without using significantly complex neural networks. 200 patients, including nasopharyngeal, lung, rectum and breast cancer cases, were collected and applied to train the proposed network. Additional 47 patients were randomly selected to evaluate the accuracy of the proposed method through comparing dose distributions, dose volume histograms (DVH) and clinical indices with the results from a treatment planning system (TPS), which was used as the ground truth in this study.

**Results:** The proposed deep learning based dose calculation algorithm achieved good predictive performance. For 47 tested patients, the average per-voxel bias of the deep learning calculated value and standard deviation (normalized to the prescription), relative to the TPS calculation, is 0.17%±2.28%. The average deep learning calculated values and standard deviations for relevant clinical indices were compared with the TPS calculated results and the t-test p-values demonstrated the consistency between them.

**Conclusions:** In this study we developed a new deep learning based dose calculation method. This approach was evaluated by the clinical cases with different sites. Our results demonstrated its feasibility and reliability and indicated its great potential to improve the efficiency and accuracy of radiation dose calculation for different treatment modalities.

## 1. INTRODUCTION

In modern radiotherapy many new dose delivery techniques, such as intensity-modulated radiation therapy (IMRT), volumetric modulated arc therapy (VMAT), stereotactic body radiation therapy (SBRT), and stereotactic radiosurgery (SRS) have been widely used in clinical practice [1]. These techniques require more accurate dose calculation for the situations with irregular small intensity-modulated beams in heterogeneous human body. In these new treatment modalities, an inverse planning is typically employed to search optimal treatment plans, which is a calculation intensive dose optimization process. Therefore, an accurate and efficient dose calculation algorithm is an important prerequisite to ensure the success of modern radiation therapy.

Although the demand for high performance dose calculation continuously increases in clinical practice, accurate and efficient dose calculation in an inhomogeneous medium such as human body is a complicated task. To date, the most accurate algorithm for dose calculation is the Monte Carlo method [2,3], which uses photon and electron transport physics to calculate the trajectories of individual particles and thus the pattern of dose deposition. However, it requires the greatest computer processing time due to the summing of the energy deposition of each individual particle in building the dose distribution. Apart from the Monte Carlo simulation, all other commonly used dose calculation algorithms can be categorized into two groups: (1) correction-based algorithms, widely used in conventional radiation therapy, which employ semi-empirical approaches to account for tissue heterogeneity and surface curvature based on measured dose distributions in water [4,5]. These methods do not perform an accurate dose calculation in patients and are rarely used now; (2) model-based algorithms, based on convolution/superposition (C/S) techniques [6-9], which predict patient dose distributions from primary particle fluence and a dose kernel. It achieves the dose calculation accuracy close to the results of Monte Carlo simulation with less time. Several variations of the C/S algorithms are implemented in commercial treatment planning system (TPS), for example, Pinnacle (Philips, Inc.) used the collapsed cone convolution (CCC) method [6], Eclipse TPS (Varian Medical Systems) used the anisotropic analytical algorithm (AAA) method [10,11] which is based on the pencil beam convolution (PBC) technique [7].

Except the Monte Carlo method, all other methods mentioned above make different degrees of approximation and simplification which leads to faster calculation speed with loss of dose calculation accuracy, thus unsatisfactory in some clinical situations. Furthermore, conventional dose calculation algorithms are limited to CT images because they perform calculations based on physics principles relying on electron densities of the medium. With the advent of MR (Magnetic Resonance) accelerators [12], a new dose calculation technology needs to be explored to break through the above limitation and calculate the dose based on the MR images directly.

In this study, we proposed a feasibility study on a new dose calculation algorithm based on the deep learning method. A voxel traversal algorithm was applied to convert a two dimensional (2D) beam fluence map to a three dimensional (3D) volume. A deep neural network was established to correlate this fluence map converted 3D volume (FMCV) with the 3D dose distribution. The performance of this dose calculation framework has been evaluated by a comprehensive dataset with different disease sites.

## 2. METHODS AND MATERIALS

We first generated FMCV, based on the beam angle and isocenter position, from a 2D beam fluence map using the voxel traversal algorithm [13]. A 3D residual network (ResNet) [14] based framework is trained to correlate the FMCV to the 3D dose distribution. The FMCV associated with the patient CT volume are the input data for the network. The output of the network is the 3D dose distribution from the given fluence map. We built the relationship between the fluence map and the 3D dose distribution, with assistance of the FMCV, using deep learning neural network. In other words, we implemented a new deep learning based dose calculation algorithm for radiotherapy. A flowchart of this proposed technique is shown in Fig. 1.

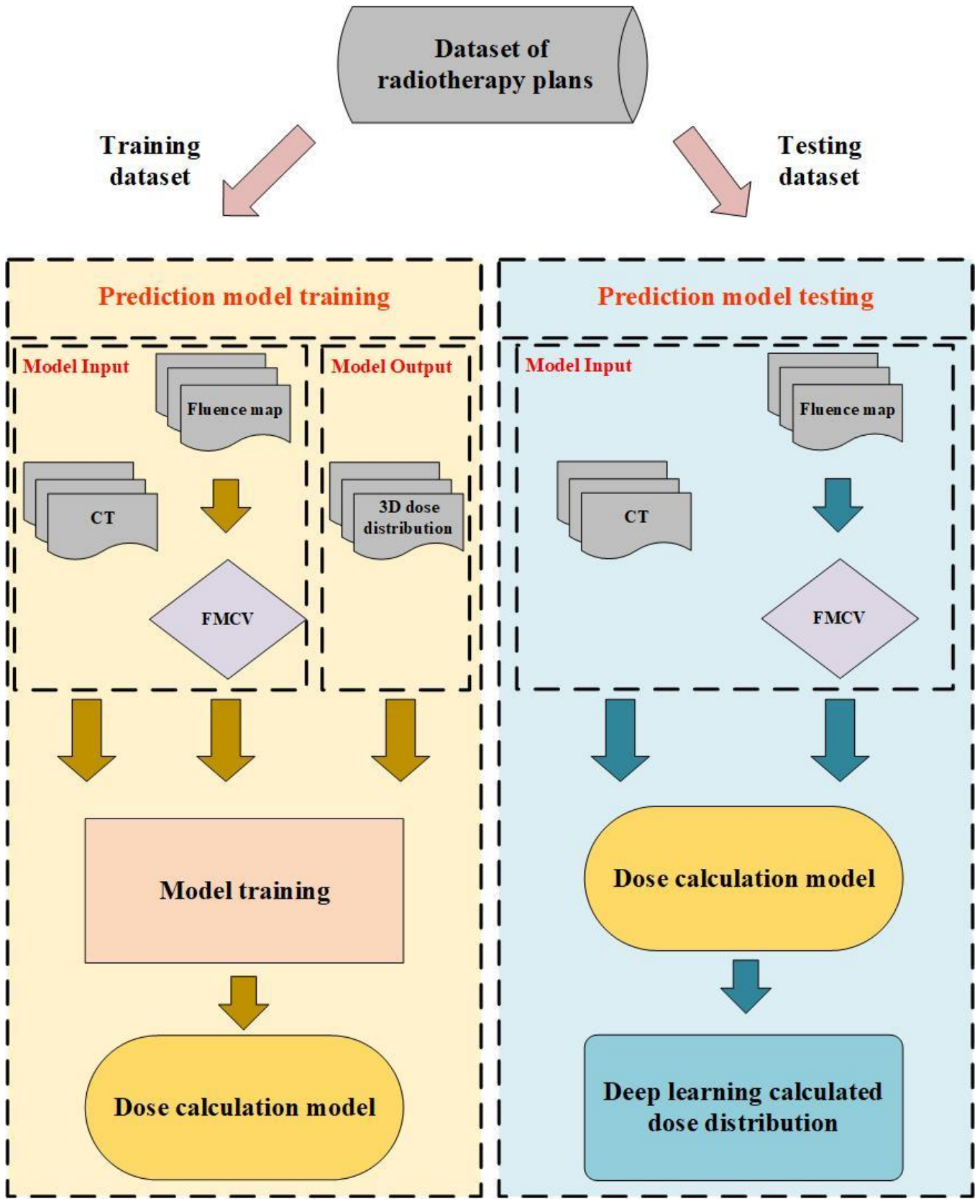

Fig 1. The flowchart of this study.

## 2.A. Patient database

The database consisted of fluence maps and CT images from 267 IMRT treated patients with different disease sites, including nasopharyngeal, lung, rectum and breast cancer cases. These data were further randomly divided into three subsets, including 200 training sets, 20 validation sets and 47 testing sets. With average 10 beams for each patient, the network uses each individual beam as the input and has ~2000 training samples extracted from Pinnacle[3] (Philips Radiation Oncology Systems, Fitchburg, WI, USA) TPS and preprocessed by our in-house DICOM processing program. The original TPS calculated dose, CT images and FMCV for all patients were resampled to a voxel resolution of $5 \times 5 \times 5$ $mm^3$ to adapt to a limited GPU memory of 12 GB.

## 2.B. Fluence map converted 3D volume

It is noted that establishing the direct relationship between the 2D fluence map with the 3D dose distribution, using neural network, is considerably difficult since the input and output of the network have different dimensions. An in-house developed algorithm based on the voxel traversal method was implemented to convert a 2D fluence map into a 3D volume. A widely used ray traversal algorithm is the three-dimensional digital differential analyzer (3D-DDA) algorithm [13]. As an efficient voxel space traversal method, it has been adopted by a number of clinical dose calculation software packages. In this study, we used 3D-DDA algorithm to iterate through the voxels along the path, in beam's eye view, connecting the ray source point and each point on the fluence map. For each penetrated voxel, a number, which is the pixel value of the connected point on the fluence map divided by the squared distance from the ray source to this penetrated voxel, was assigned as its voxel value. We emphasize that the proposed FMCV is rather a simple technique to convert 2D fluence map into 3D volume. And our results indicate that other dose calculation related features, including tissue inhomogeneity, attenuation and dose deposition in the patient anatomy can be learned from network training. Fig. 2 illustrates fluence maps, with different beam angles 120 and 240, and their corresponding one slice image through the isocenter of FMCV volumes.

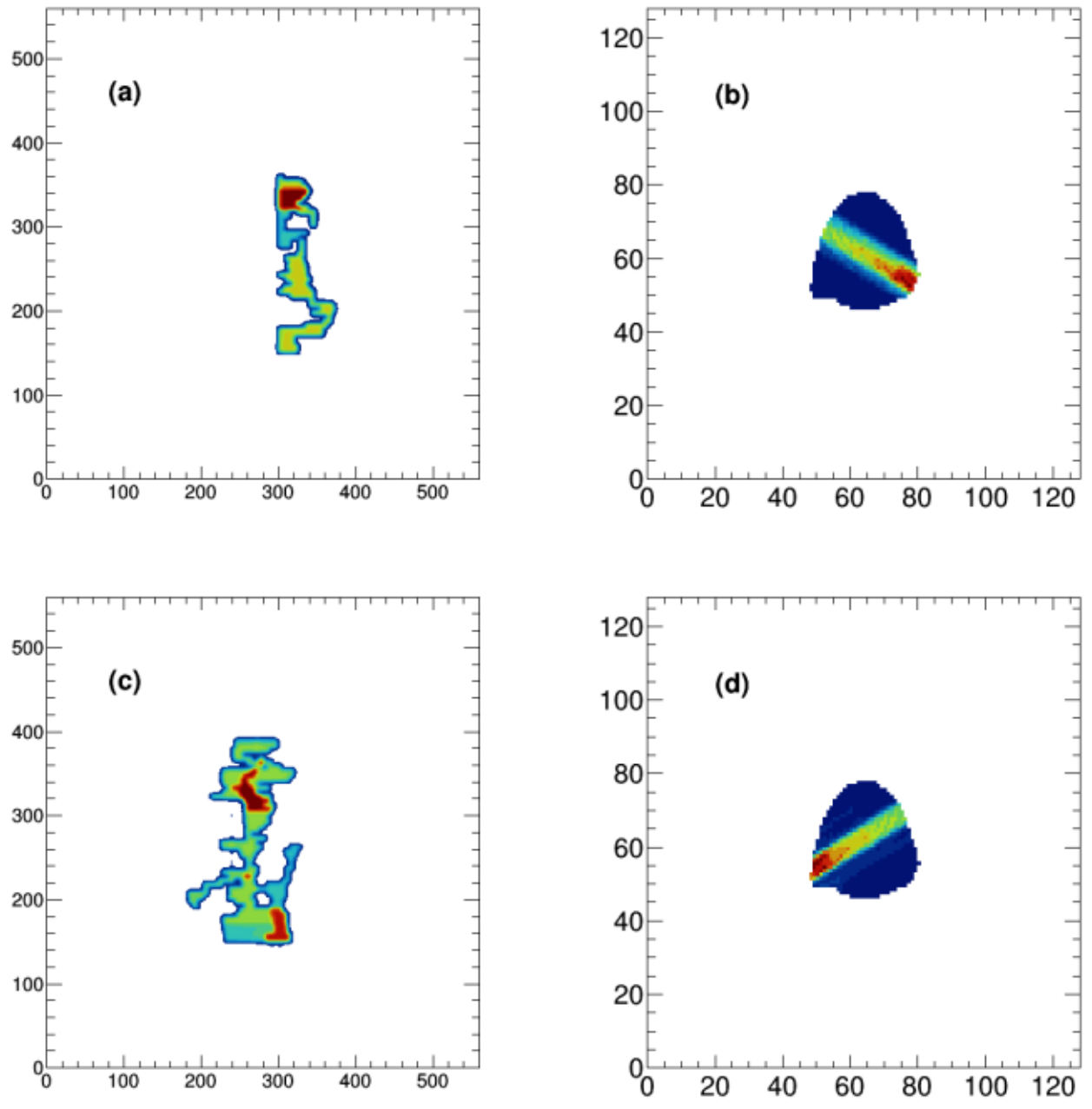

Fig. 2. Two fluence maps (a, c), with beam angles 120 and 240, and their corresponding one slice image through the isocenter of FMCV volumes (b, d).

### 2.C. Model architecture

The architecture of the deep learning model is illustrated in Fig. 3. Our group has previously used a similar architecture to generate dose distribution prediction for a given patient anatomy [15]. In this study, concerning the calculation errors due to the absence of enough superior and inferior CT information, we developed a 3D model constructed by 3D convolutional layers to replace the published 2D model.

As can be seen in Fig. 3, the 3D CT volume and FMCV were treated as different input channel in this model. The output dose distribution can be obtained by first downsampling the input data followed by the upsampling associated with the long skip connections. The skip connection was applied to combat the gradient vanishing and gave us the ability to train deeper networks [14-16]. The downsampling and upsampling parts consisted of different combination of stacked building blocks of Identity block, Convolution block and Transposed convolution block (shown in Fig. 3) with 1x1 and 3x3 kernels. The downsampling process was achieved by one convolutional layer in the Convolution block with stride setting to 2 while the upsampling process was implemented by setting the stride to 2 for one transposed convolution layer in the Transposed convolution block. To maintain a reasonable GPU memory usage, the maximum number of filters for the convolutional layer in the network was set to 512. To prevent overfitting, the dropout technique, which is a regularization technique that randomly removes or drops out some neurons by a dropout rate between 0 and 1 at each training update, was employed. It helps to reduce codependency amongst the neurons and thus prevent the overfitting. Also, the data augmentation technique, randomly shifting the 3D CT volume, FMCV and 3D dose distribution in X or Y axis, was implemented to further prevent the overfitting.

ReLU activation function was applied right after each convolution and transposed convolution layer [17]. The Adam optimization was used to minimize the mean squared error (MSE) loss function with a mini-batch size of 2. The learning rate was set to 0.0001, and about 200 training epochs were taken until the validation loss did not significantly decrease during training. We implemented this predictive model in Keras [18], which is a high-level open-source deep learning library written in Python and capable of running on top of TensorFlow. We initialized the weights randomly and trained the network from scratch, which took about 3 weeks of computation time, using NVIDIA 2080Ti GPU with 12 GB memory.

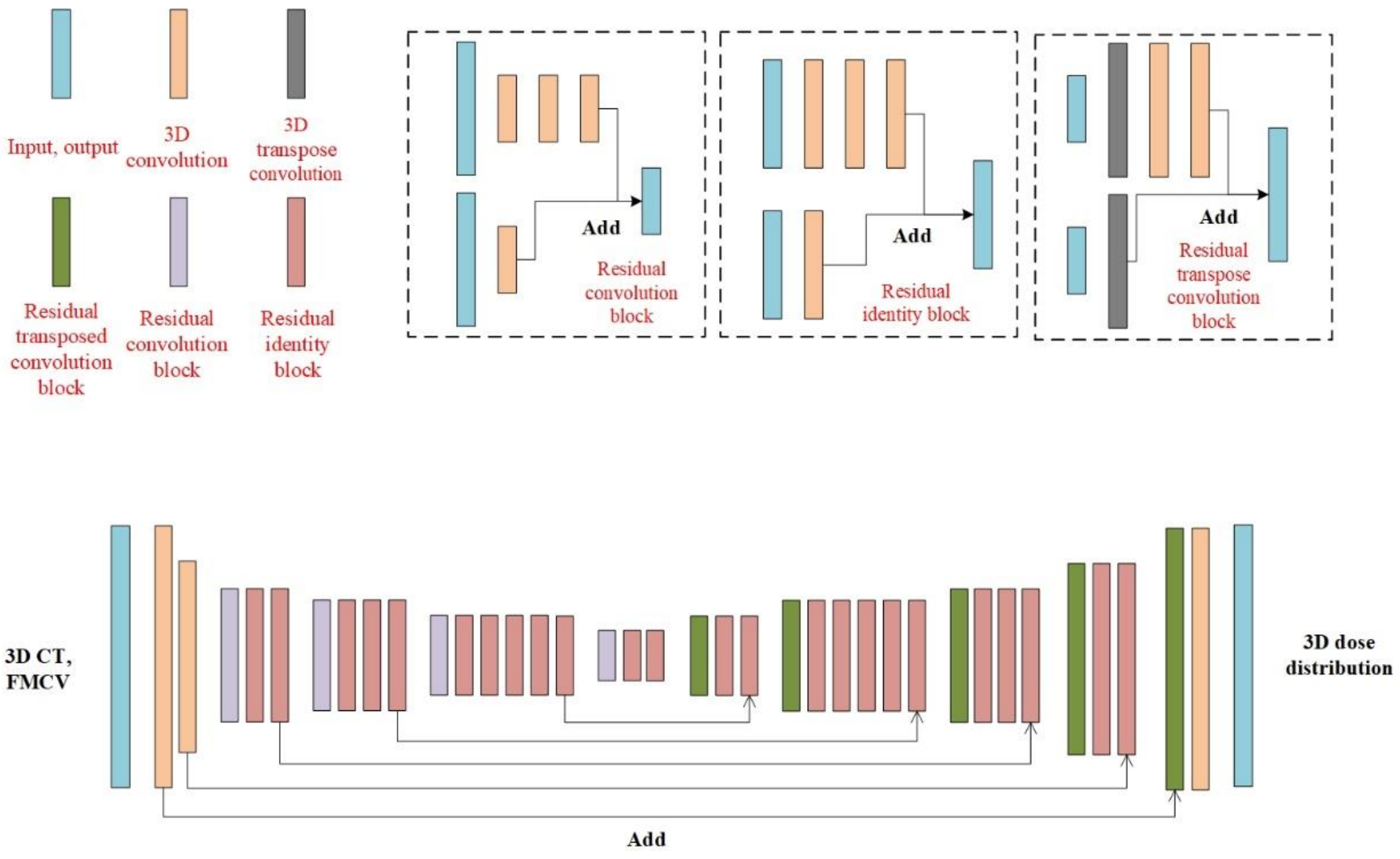

Fig. 3. The architecture of the proposed deep learning model.

### 2.D. Model performance

To evaluate the model's performance and stability, 47 cases, including nasopharyngeal, lung, rectum and breast cancer patients, were randomly selected as independent testing data. For each fluence map, we calculated the corresponding dose distribution by applying the trained model. The patient's entire deep learning model calculated (DL calculated) dose distribution was obtained by summing doses from all treatment beams. The DL calculated dose volume histograms (DVH) for targets and organs at risk (OARs), and their relative clinical indices were further compared with the values computed from TPS.

## 3. RESULTS

The proposed deep learning based dose calculation method achieved acceptable performance in terms of the dose distribution and DVH comparison. The dose distributions for each beam in 47 new patients were calculated by applying the trained model and the average calculation time is less than several seconds. The patients' overall dose distributions were obtained by summing doses from all individual beams in the plan. Fig. 4 and Fig. 5 show the TPS calculated and DL calculated dose distributions, as well as pixel-wise dose differences in two axial planes for a nasopharyngeal and lung case, respectively. The similar results for breast and rectum patient are shown in the Appendix. For all the 47 tested cases, the average per-voxel bias of the DL calculated dose and standard deviation (normalized to the prescription) relative to the TPS calculated dose is shown in Fig. 6, categorized by disease sites. As we can see, all the average per-voxel bias of the DL calculated dose is within 3%.

Fig. 7 shows the comparison between the TPS and DL calculated DVHs of OARs and targets for a nasopharyngeal, lung, breast and rectum patient. The statistical results of clinical indices for OARs and targets, calculating from all the testing nasopharyngeal, lung, breast and rectum patients, are reported in Table 1, respectively. The average values and their standard deviations for relevant clinical indices were compared and an agreement between the TPS and DL calculated results can be seen from Table 1. The t-test p-values, which further test the consistency of the TPS and DL calculated results from the statistical point of view, in the last column of Table 1 also verify the accuracy of the DL calculated results.

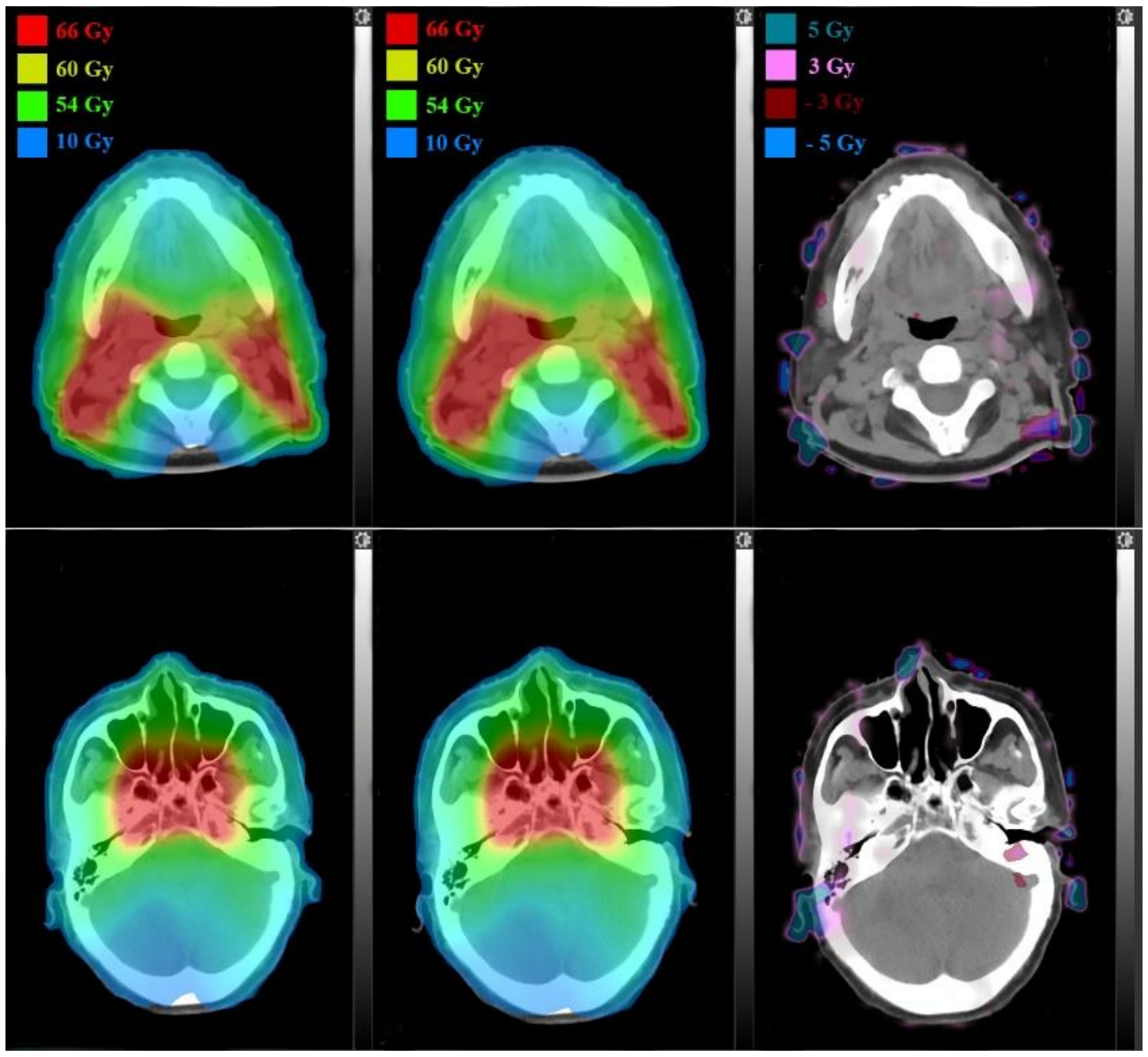

Fig. 4. Dose distributions comparison in two axial planes for a nasopharyngeal patient: TPS calculated (left), DL calculated (middle) and pixel-wise differences (right).

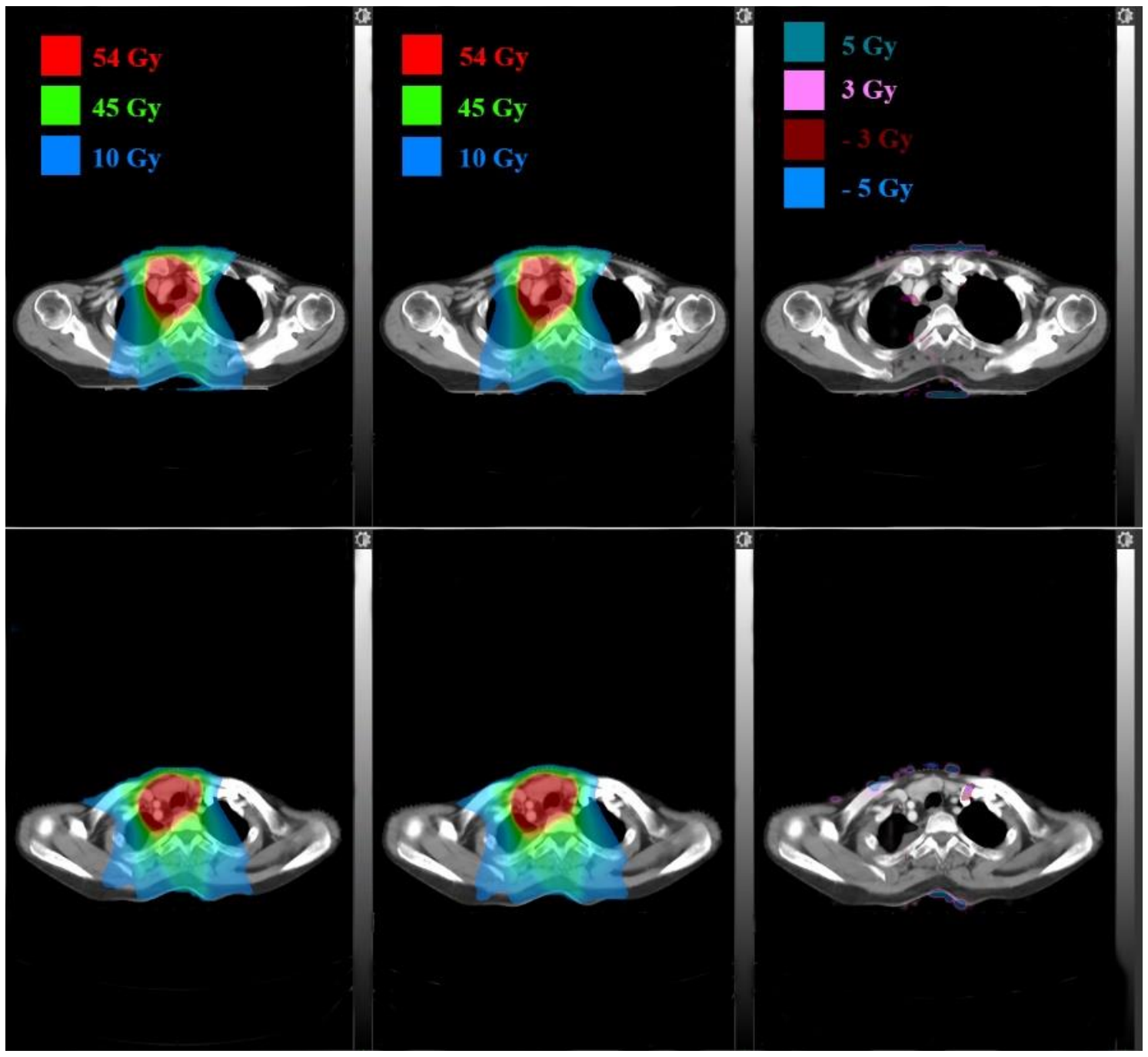

Fig. 5. Dose distributions comparison in two axial planes for a lung patient: TPS calculated (left), DL calculated (middle) and pixel-wise differences (right).

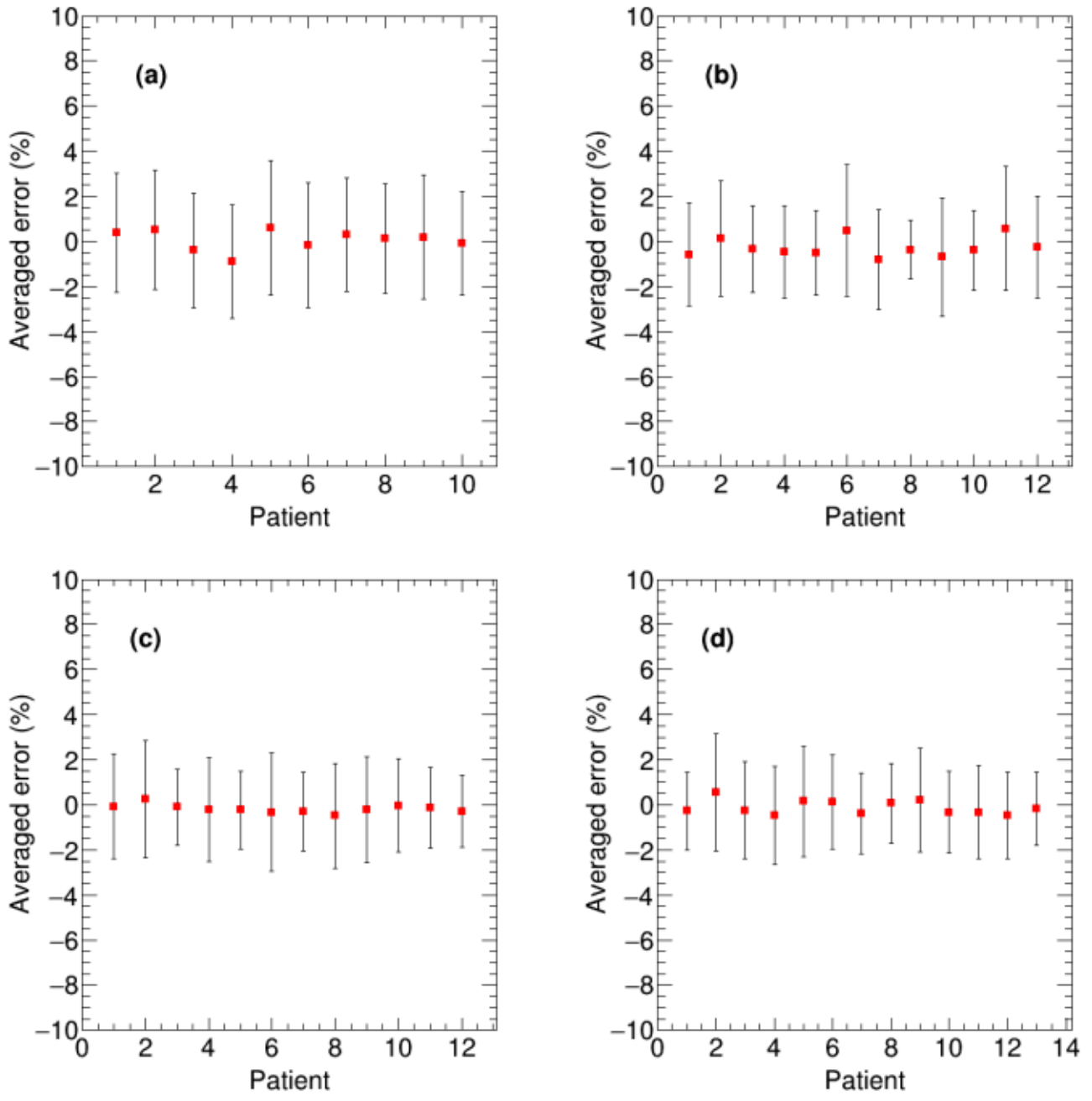

Fig. 6. The average per-voxel bias of the DL calculated dose and standard deviation (normalized to the prescription) relative to the TPS calculated dose for all 47 test cases categorized by disease site: nasopharyngeal (a), lung (b), breast (c) and rectum (d).

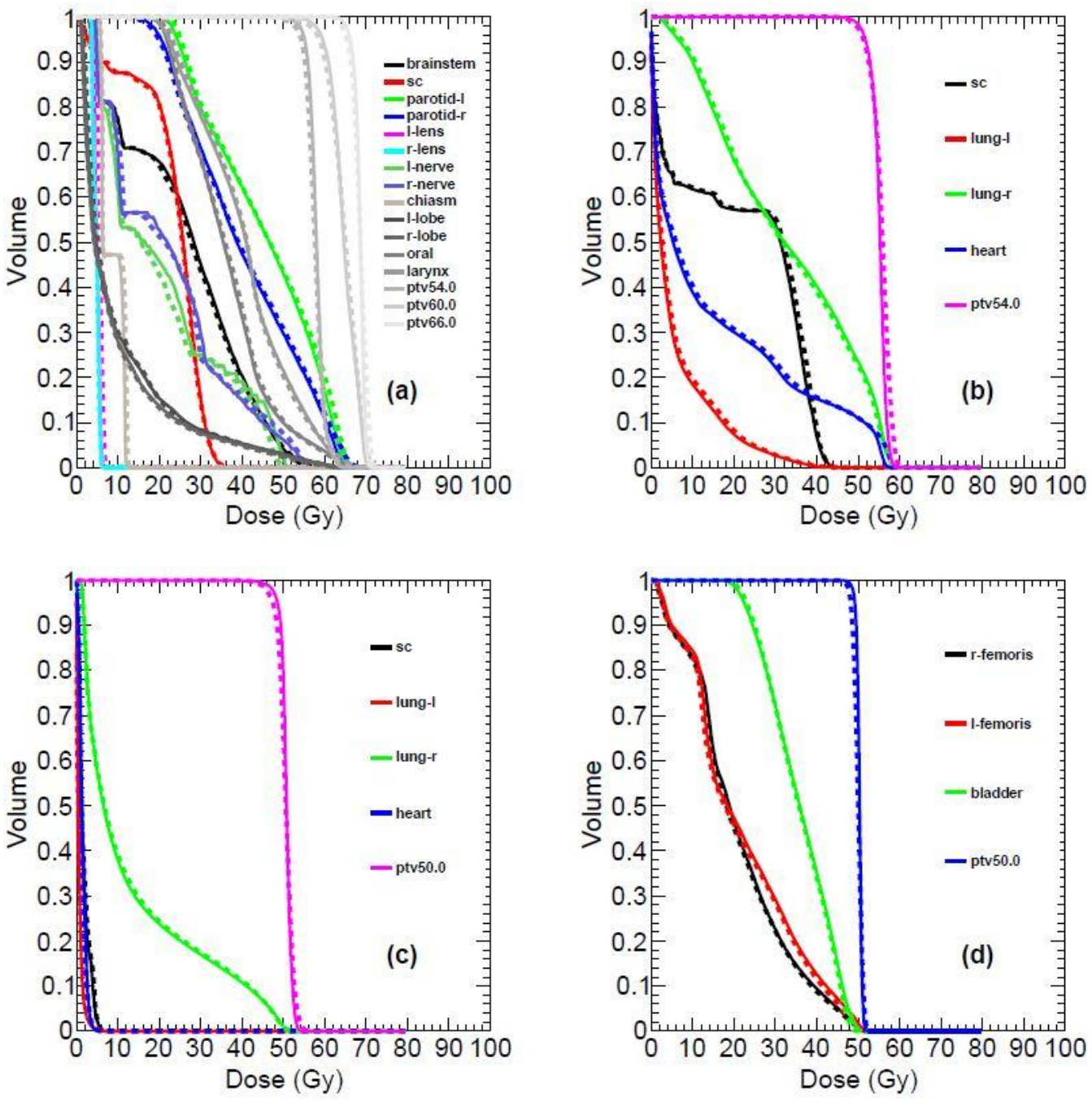

Fig. 7. Comparison between the TPS calculated (solid line) and DL calculated (dashed line) OARs and target DHVs for different disease sites: nasopharyngeal (a), lung (b), breast (c) and rectum (d).

Table 1. The average value and standard deviation (mean ± SD) of relevant clinical indices for all the testing patients.

| Structure | Clinical indices (Gy) | TPS calculation | DL calculation | t-test p-value |
|---|---|---|---|---|
| **Nasopharyngeal** | | | | |
| Brainstem | $D_{max}$ | 53.45 ± 1.85 | 55.50 ± 2.12 | 0.05 |
| Left parotid gland | $D_{mean}$ | 41.76 ± 3.21 | 42.56 ± 3.01 | 0.59 |
| | $V_{30}$ (%) | 68.42 ± 7.48 | 71.70 ± 7.06 | 0.35 |
| Right parotid gland | $D_{mean}$ | 38.07 ± 3.73 | 38.18 ± 3.38 | 0.95 |
| | $V_{30}$ (%) | 60.67 ± 10.51 | 62.20 ± 8.94 | 0.74 |
| Spincal cord | $D_{max}$ | 40.55 ± 1.80 | 42.45 ± 2.27 | 0.06 |
| Left len | $D_{max}$ | 6.60 ± 1.26 | 7.30 ± 1.14 | 0.23 |
| Right len | $D_{max}$ | 6.50 ± 0.87 | 7.00 ± 0.89 | 0.24 |
| Left nerve | $D_{max}$ | 52.95 ± 7.19 | 52.55 ± 7.72 | 0.91 |
| Right nerve | $D_{max}$ | 56.65 ± 8.65 | 52.45 ± 9.11 | 0.96 |
| Chiasm | $D_{max}$ | 37.30 ± 18.62 | 37.45 ± 17.53 | 0.99 |
| Left temporal lobe | $D_{max}$ | 67.85 ± 4.95 | 67.65 ± 4.67 | 0.93 |
| Right temporal lobe | $D_{max}$ | 67.80 ± 5.01 | 68.25 ± 5.36 | 0.86 |
| Oral cavity | $D_{mean}$ | 36.02 ± 4.74 | 36.55 ± 4.47 | 0.81 |
| Larynx | $D_{mean}$ | 36.50 ± 3.44 | 36.41 ± 3.44 | 0.96 |
| **Lung** | | | | |
| Heart | $D_{mean}$ | 11.99 ± 7.71 | 12.17 ± 7.61 | 0.96 |
| Spinal cord | $D_{max}$ | 40.79 ± 2.59 | 42.71 ± 2.79 | 0.11 |
| **Breast** | | | | |
| Left lung | $V_{20}$ (%) | 7.81 ± 7.65 | 8.00 ± 7.90 | 0.97 |
| Right lung | $V_{20}$ (%) | 11.17 ± 10.30 | 11.32 ± 10.45 | 0.98 |
| Heart | $D_{mean}$ | 2.44 ± 1.70 | 2.50 ± 1.69 | 0.94 |
| **Rectum** | | | | |
| Left femoris | $V_{30}$ (%) | 20.65 ± 12.39 | 21.40 ± 13.77 | 0.89 |
| Right femoris | $V_{30}$ (%) | 17.11 ± 11.51 | 17.47 ± 11.92 | 0.94 |
| Bladder | $V_{30}$ (%) | 64.56 ± 23.26 | 66.40 ± 23.79 | 0.85 |

## 4. DISCUSSION

Deep learning is widely used for achievable dose prediction and OARs or targets segmentation in radiation therapy research [15, 19-24]. Recently, Dong and Xing pioneered the use of deep neural network for dose calculation [22, 25]. Thereafter, there are few relevant studies on the introduction of deep learning for dose calculation [26-28]. In contrast to previously published studies, in this work we proposed a new simple technique to convert 2D fluence map into 3D volume that can be used to build the dose calculation deep learning model. The model performance was evaluated in a more comprehensive dataset, including four different disease sites and multiple beam angles. As can be seen in Fig. 4-7 and Table 1, our preliminary results show that the proposed method is a promising new technique for dose calculation and the results are within clinical acceptable limit.

One of the key technologies in this study is the implementation of the FMCV, instead of 2D fluence map, as the model input. This avoids the adoption of a more complex neural network and makes the model training much easier. We know that the 3D dose depends on the isocenter position for a given fluence map. As stated in Section 2.B, the path connecting the source point and the point on the fluence map is affected by the geometry position of the fluence map. The generation procedure of the FMCV takes naturally the isocenter position into account and different isocenter position yields different FMCV even if the fluence map is the same. We emphasize that the inverse square law is considered in the FMCV generation process, other dose calculation related features, such as tissue inhomogeneity, attenuation and dose deposition in the patient anatomy can be learned from network training.

Our results demonstrate that the proposed model has high dose calculation accuracy with great efficiency. For 47 tested patients, the average per-voxel bias of the DL calculated dose and standard deviation (normalized to the prescription) is 0.17% ± 2.28% relative to the TPS calculated results. As noted in Fig. 4-5, the dose differences become slightly larger in the boundary region. This is on account of large variation of material density in the boundary region, which leads to the higher uncertainty and inconsistent clinical delivered dose calculation results using the CCC algorithm in Pinnacle[3] system. In this study we used the 3D doses calculated with the commercial TPS dose algorithm (CCC algorithm in Pinnacle) as input to train the model and also used as the benchmarks to evaluate the dose accuracy for our proposed mothed. This is acceptable to study the feasibility of our proposed method. However, in order to obtain more consistency with actually irradiated dose in patients and implement the proposed technique in clinical use, it is better to use the dose distributions calculated with more accurate dose algorithms, such as Monte Carlo algorithm, as the input to train the model. In this way, we believe that the proposed method is able to provide dose calculation accuracy similar to the Monte Carlo simulation with higher efficiency. In addition, although the dose calculation for photon beams was tested in this study, the proposed method is general enough and can be easily applied to calculate dose distributions for other particles, such as proton and electron beams.

The proposed method is capable of calculating the dose distribution given the CT and fluence map. In the model, the relationship between input and output dose distribution, at the pixel level, is established. In this context, we can establish a model applicable to other images besides CT images, such as MR images. Therefore, the proposed deep learning-based dose calculation method can take the MR images as the input and train a model dedicated to perform dose calculation on MR images. This new dose calculation algorithm has great potential to improve dose calculation accuracy and efficiency of the MR image-based dose calculation and facilitate the clinical procedure for the MR accelerator-based treatment.

## 5. CONCLUSION

In this study we developed a new deep learning based method to perform accurate and efficient dose calculation. This approach was evaluated by the clinical cases with different sites. Our results demonstrated its feasibility and reliability and indicated its great potential to improve the efficiency and accuracy of radiation dose calculation for different treatment modalities.

## ACKNOWLEDGEMENTS

This work was partially supported by NIH (R01CA227713 and R01CA223667) and a Google Faculty Research Award (LX) and National Natural Science Foundation of China (No. 11805039) (JF).

## DISCLOSURE of CONFLICTS of INTEREST

None

## Appendix

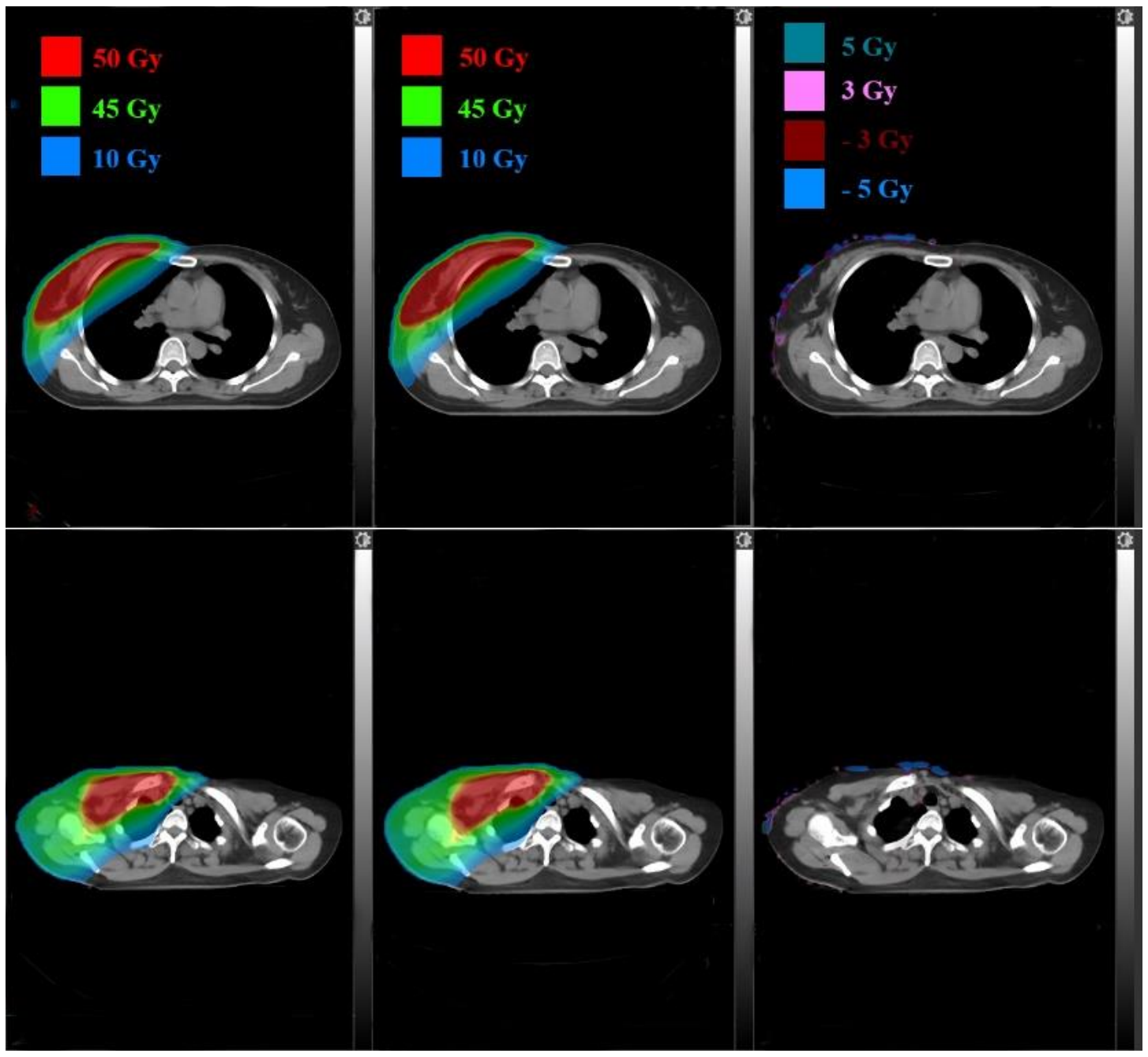

Fig. 8. Dose distributions comparison in two axial planes for a breast patient: TPS calculated (left), DL calculated (middle) and pixel-wise differences (right).

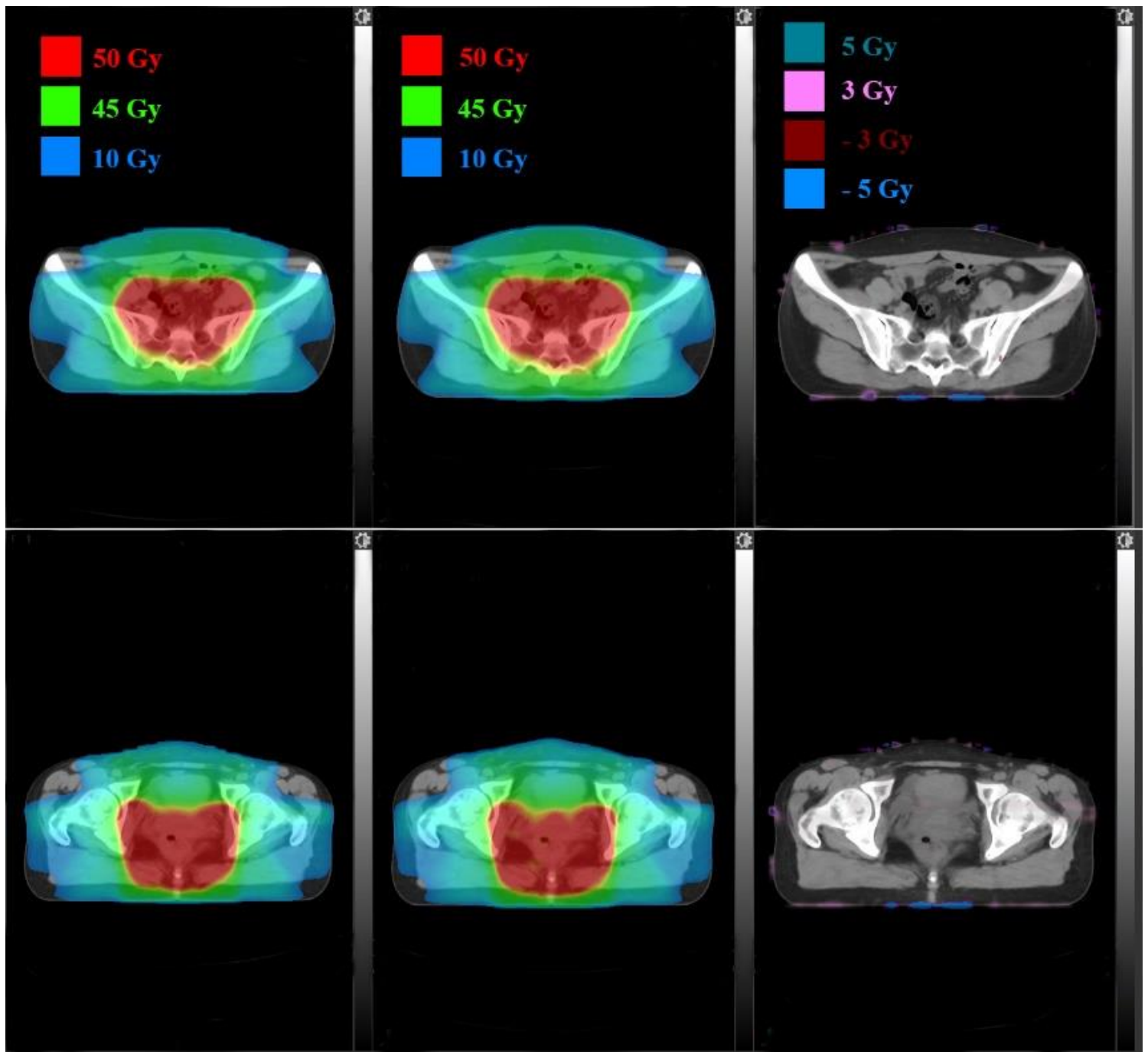

Fig. 9. Dose distributions comparison in two axial planes for a rectum patient: TPS calculated (left), DL calculated (middle) and pixel-wise differences (right).